\documentclass[twocolumn,showpacs,preprintnumbers,amsmath,amssymb,superscriptaddress,groupedaddress,prb]{revtex4-1}

\usepackage{epsfig}                                                            
\usepackage{graphicx}
\usepackage{dcolumn}
\usepackage{color}
\usepackage{bm}
\usepackage{subfigure} 
\usepackage{epsfig} 

\begin{document}

\title{\bf Particle size dependence of magnetization and non-centrosymmetry in nanoscale BiFeO$_3$}

\author{Sudipta Goswami} \affiliation{Nanostructured Materials Division, Central Glass and Ceramic Research Institute, CSIR, Kolkata 700032, India}
\author{Dipten Bhattacharya} 
\email{dipten@cgcri.res.in}\affiliation{Nanostructured Materials Division, Central Glass and Ceramic Research Institute, CSIR, Kolkata 700032, India}
\author{P. Choudhury} \affiliation{Nanostructured Materials Division, Central Glass and Ceramic Research Institute, CSIR, Kolkata 700032, India}

\date{\today}

\begin{abstract}

The saturation magnetization (M$_S$), antiferromagnetic transition point (T$_N$), and the off-center displacements of Bi and Fe ions have been measured as a function of particle size in nanoscale BiFeO$_3$. T$_N$ decreases down to $\sim$550 K for particles of size $\sim$5 nm from $\sim$653 K in bulk while M$_S$ rises by more than an order of magnitude. Analysis of crystallographic structure from Rietveld refinement of x-ray diffraction patterns shows significant rise in off-center displacements of Bi ($\delta_{Bi}$) and Fe ($\delta_{Fe}$) ions within a unit cell with the decrease in particle size. The net unit-cell polarization P$_S$ too, is found to be larger in nanoscale regime.

\end{abstract}
\maketitle 

\section{Introduction}
Extensive research work has been done on BiFeO$_3$ over the decades because of its magnetoelectric multiferroicity at room temperature.\cite{Schmid} The linear magnetoelectric coupling originates from striction mediated interaction between polar and magnetic domains.\cite{Lee} The bulk BiFeO$_3$ is G-type antiferromagnetic with a spiral of long period (~62 nm). Therefore, the magnetostriction is weak which, in turn, gives rise to weak coupling. The electric polarization is also small ($\sim$5-6 $\mu$C/cm$^2$) in bulk.\cite{Teague} It improves ($\sim$55 $\mu$C/cm$^2$) in thin films due to the strain between substrate and film.\cite{Wang} Large polarization and magnetization will render BiFeO$_3$ quite useful as a single phase multiferroic compound.  
In this paper, we show that it is possible to achieve both the aspects - large polarization and magnetization - in nanoscale BiFeO$_3$ ($\sim$5-50 nm). We found that intrinsic ferromagnetism develops in nanoscale BiFeO$_3$ with rise in saturation magnetization M$_S$, decrease in antiferromagnetic transition point T$_N$, and increase in coercive field H$_C$. We also found that the unit-cell off-center displacements of Bi and Fe ions - which render BiFeO$_3$ ferroelectric - increases monotonically with the decrease in particle size. The net unit-cell polarization P$_S$ too, is found be higher in nano-sized particles.

\section{Experiments}
The experiments were carried out on both bulk and nano-sized particles of BiFeO$_3$. The nanoparticles of BiFeO$_3$ have been synthesized by sonochemical process where coprecipitation takes place from mixed aqueous solution of metal nitrates within a suitable medium in presence of ultrasonic vibration. The precipitate was washed in alcohol and dried. Finally, the product was calcined at 350$^o$-450$^o$C for 2-6h. The sonochemical process was shown to yield nanosized particles of various oxide compounds in the past. We have separated out finer particles by a centrifuge running at different speeds - 10000, 12000, and 15000 rpm. The variation of heat-treatment temperature and time also yields particles of different sizes. In order to synthesize the bulk sample, the calcined powder was compacted in the form of pellets and sintered at ~830$^o$C for 5-10h. The average grain size in the sintered pellet was $\geq$0.1$\mu$m.

The samples have been characterized by studying the x-ray diffraction (XRD) patterns at room temperature and from the transmission electron microscopy (TEM) and high resolution TEM. The electrical resistivity and magnetic measurements were carried out across a temperature range 300-900 K. The resistivity of the particles was measured by depositing a coating of the particles on an alumina substrate with gold electrodes printed on it. The XRD patterns were refined by Fullprof (ver 2.3) to determine the structural parameters such as space group, lattice parameters, crystallite size, strain, atom positions, bond lengths and angles etc. The off-center displacements of Bi$^{3+}$ ($\delta_{Bi}$) and Fe$^{3+}$ ($\delta_{Fe}$) ions have also been calculated as a function of particle size. 

\section{Results and Discussion}
The average size of the nanoscale BiFeO$_3$ particles varies from $\sim5$ to $\sim$50 nm. In Fig. 1, we show a representative TEM photograph. The average size was estimated by image analyzer software Image-J. It is compared with the crystallite size estimated from the Rietveld refinement of x-ray diffraction patterns. Both the results corroborate each other.

\begin{figure}[!h]
\centering
\includegraphics[scale=.37]{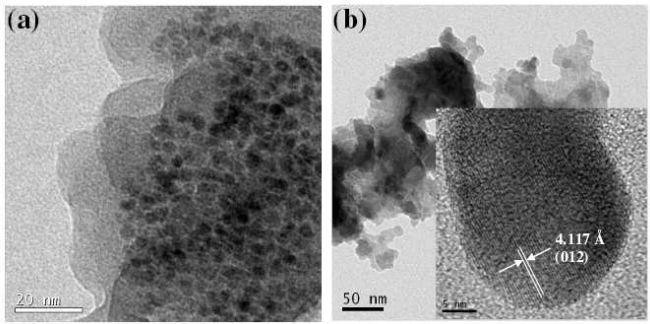}
\caption{TEM photographs for (a) finer and (b) coarser BiFeO$_3$ particles; inset shows the high resolution TEM photograph of a single crystalline nanoparticle.}
\end{figure}

\begin{table*}[!ht]
\caption{List of lattice parameters, atom positions, bond length, angle, and microstrain at room temperature. }

\begin{tabular}{p{0.5in}p{0.6in}p{0.5in}p{0.5in}p{0.4in}p{0.4in}p{0.5in}p{0.5in}p{0.5in}p{0.5in}p{0.5in}p{0.7in}p{0.5in}} \hline \hline
Samples & Lattice \newline parameters \newline $(\AA)$ & Atoms & Positions & x & y & z & Bonds & length & Bonds & angle & microstrain & $R_{wp}$ \\ \hline 
\\
BiFeO$_3$\newline -bulk & a = 5.578 \newline c = 13.868 & Bi \newline Fe \newline O & 6a \newline 6a \newline 18b & 0.0 \newline 0.0 \newline 0.4346 & 0.0 \newline 0.0 \newline 0.0121 & 0.0 \newline 0.2198 \newline -0.0468 & Bi-O \newline Fe-O \newline Fe-O & 2.309$\AA$ \newline 1.949 $\AA$ \newline 2.118 $\AA$ & Fe-O-Fe \newline O-Bi-O & 154.05$^o$ \newline 73.88$^o$ & 0.015 & 22.6 \\  
\\
BiFeO$_3$ \newline - 25 nm & a = 5.573 \newline c = 13.849 & Bi \newline Fe \newline O & 6a \newline 6a \newline 18b & 0.0 \newline 0.0 \newline 0.4715 & 0.0 \newline 0.0 \newline 0.0119 & 0.0 \newline 0.22324 \newline -0.0622 & Bi-O \newline Fe-O \newline Fe-O & 2.130 $\AA$ \newline 1.804 $\AA$ \newline 2.269 $\AA$ & Fe-O-Fe \newline O-Bi-O & 152.76$^o$ \newline 78.93$^o$ & 0.029 & 6.68 \\ 
 \\ 
BiFeO$_3$\newline - 22 nm & a = 5.579 \newline c = 13.870 & Bi \newline Fe \newline O & 6a \newline 6a \newline 18b & 0.0 \newline 0.0 \newline 0.4455 & 0.0 \newline 0.0 \newline 0.03123 & 0.0 \newline 0.2256 \newline -0.05351 & Bi-O \newline Bi-O \newline Fe-O \newline Fe-O & 2.179 $\AA$ \newline 2.515 $\AA$ \newline 1.955 $\AA$ \newline 2.172 $\AA$ & Fe-O-Fe \newline O-Bi-O & 147.77$^o$ \newline 73.84$^o$ & 0.049 & 11.7 \\ 
 \\ 
BiFeO$_3$\newline - 19 nm & a = 5.624 \newline c = 13.672 & Bi \newline Fe \newline O & 6a \newline 6a \newline 18b & 0.0 \newline 0.0 \newline 0.3712 & 0.0 \newline 0.0 \newline -0.0898 & 0.0 \newline 0.2285 \newline 0.00216 & Bi-O \newline Bi-O \newline Fe-O \newline Fe-O & 2.381 $\AA$ \newline 2.586 $\AA$ \newline 1.941 $\AA$ \newline 2.267 $\AA$ & Fe-O-Fe \newline O-Bi-O & 140.91$^o$  \newline 85.17$^o$ & 0.105 & 15.8 \\ 
\\ \hline \hline

\end{tabular}
\end{table*}
 
The magnetic measurements were carried out in zero-field cooled mode across 300-900 K under an applied field $\sim$100 Oe. In Fig. 2a, we show the magnetization versus temperature plot for a few representative samples. Quite evident is the monotonic increase in magnetization with the decrease in particle size. This observation corroborates earlier observations by us as well as by other authors.\cite{Mazumder}$^,$\cite{Park} Interestingly, the antiferromagnetic transition point T$_N$ drops significantly with the decrease in particle size: from $\sim$653 K in bulk to $\sim$550 K in particles of average size $\sim$5 nm. Dc electrical resistance has also been measured across 300-800 K. Around T$_N$, a distinct feature in the resistance versus temperature plot could be noticed (data not shown here). T$_N$ was thus estimated both from the magnetic and electrical measurements. In Fig. 2b, the T$_N$ versus particle size plot is shown. There is a certain discrepancy between T$_N$ estimated from magnetic and electrical measurements, especially, in finer particle regime. This could result from progressive broadening of the transition zone. We have carried out Curie-Weiss fitting of the magnetization data in the paramagnetic regime (Fig. 2a inset). With the decrease in particle size, the Weiss constant $\theta$ appears to be switching from negative (signaling antiferromagnetic order) to positive (marking ferromagnetism) regime. The degree of frustration $\textit{f}$ = $\theta$/T$_N$ increases from $\sim$1.0 to $\sim$1.4 with the decrease in particle size. This could be because of enhanced geometry-driven frustration in spin ordering within near neighbors in finer particles. This has been observed in nanosized ferromagnetic particles by others as well.\cite{Guimaraes} The Curie constant C yields the magnetization to be varying within $\sim$0.108-0.452 $\mu_B$/Fe for particle size varying within 5-50 nm. For the bulk system, the magnetization is $\sim$0.02 $\mu_B$/Fe. The room temperature saturation magnetization M$_S$ and coercivity H$_C$ are plotted as a function of particle size in Fig. 2b.  The intrinsic ferromagnetism in nanoscale BiFeO$_3$ results from\cite{Mazumder} (i) incomplete spiral of magnetic order in particles of size less than $\sim$62 nm, and (ii) enhanced strain of the nano-crystals.

\begin{figure}[!h]
\begin{center}
          \subfigure[]{\includegraphics[scale=.25]{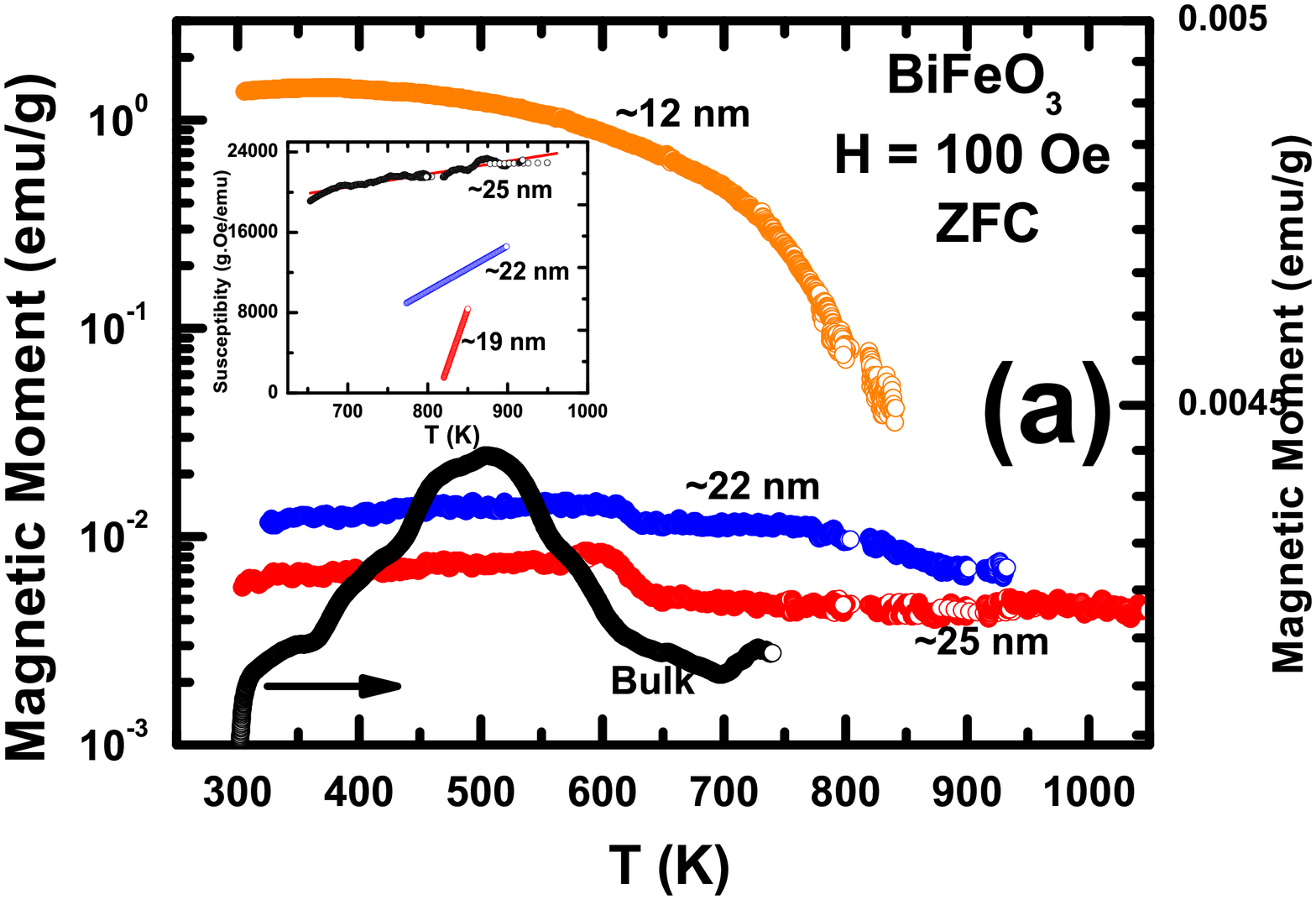}}\\
          \subfigure[]{\includegraphics[scale = 0.25]{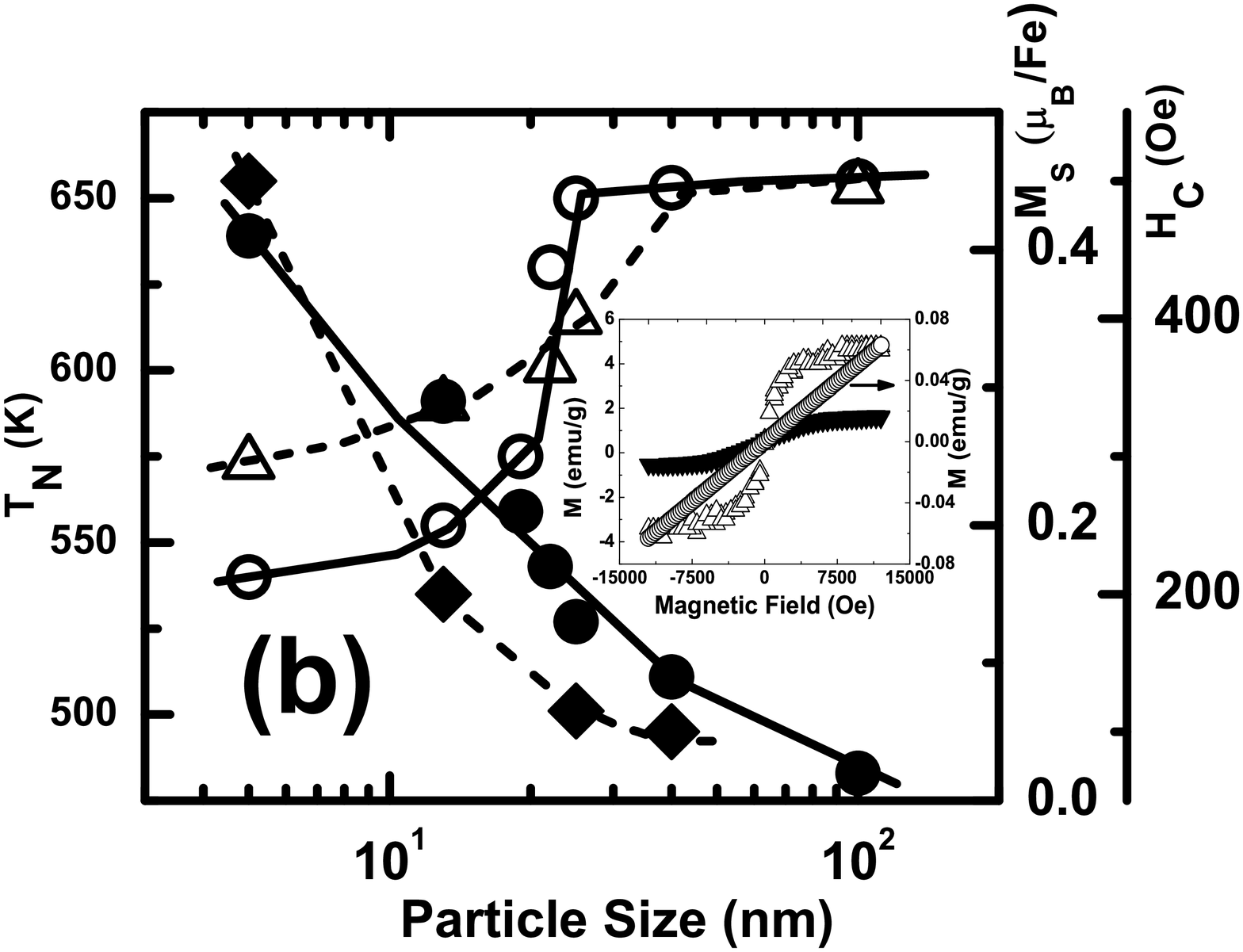}}
\end{center}
\caption{(color online) (a) Magnetic moment versus temperature plots for a few representative samples of different particle size; the magnetic transition near T$_N$ could be clearly observed in bulk and coarser particles; inset shows the Curie-Weiss plot in the paramagnetic regime; (b) variation in the magnetic transition point T$_N$ as estimated from magnetic (open circle) and electrical (up-triangle) measurements; the variation of M$_S$ (solid circle) and H$_C$ (diamond) with particle size are also shown; inset shows the room temperature hysteresis loops for bulk (open circle) and nanoparticles of size ~25 nm (up-triangle) and ~40 nm (down-triangle).}
\end{figure}

\begin{figure}[!h]
\begin{center}
          \subfigure[]{\includegraphics[scale=.25]{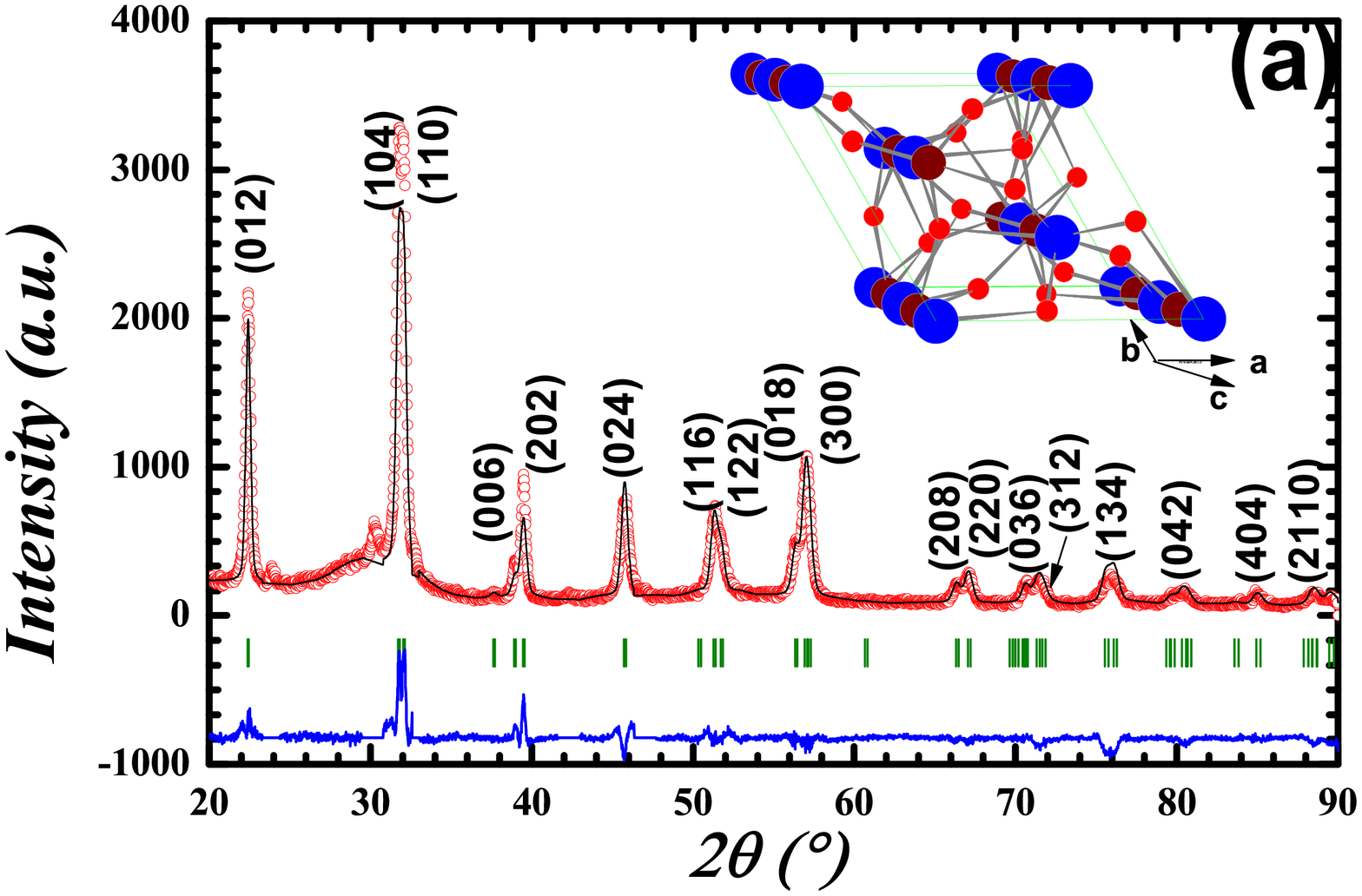}}\\
          \subfigure[]{\includegraphics[scale=.25]{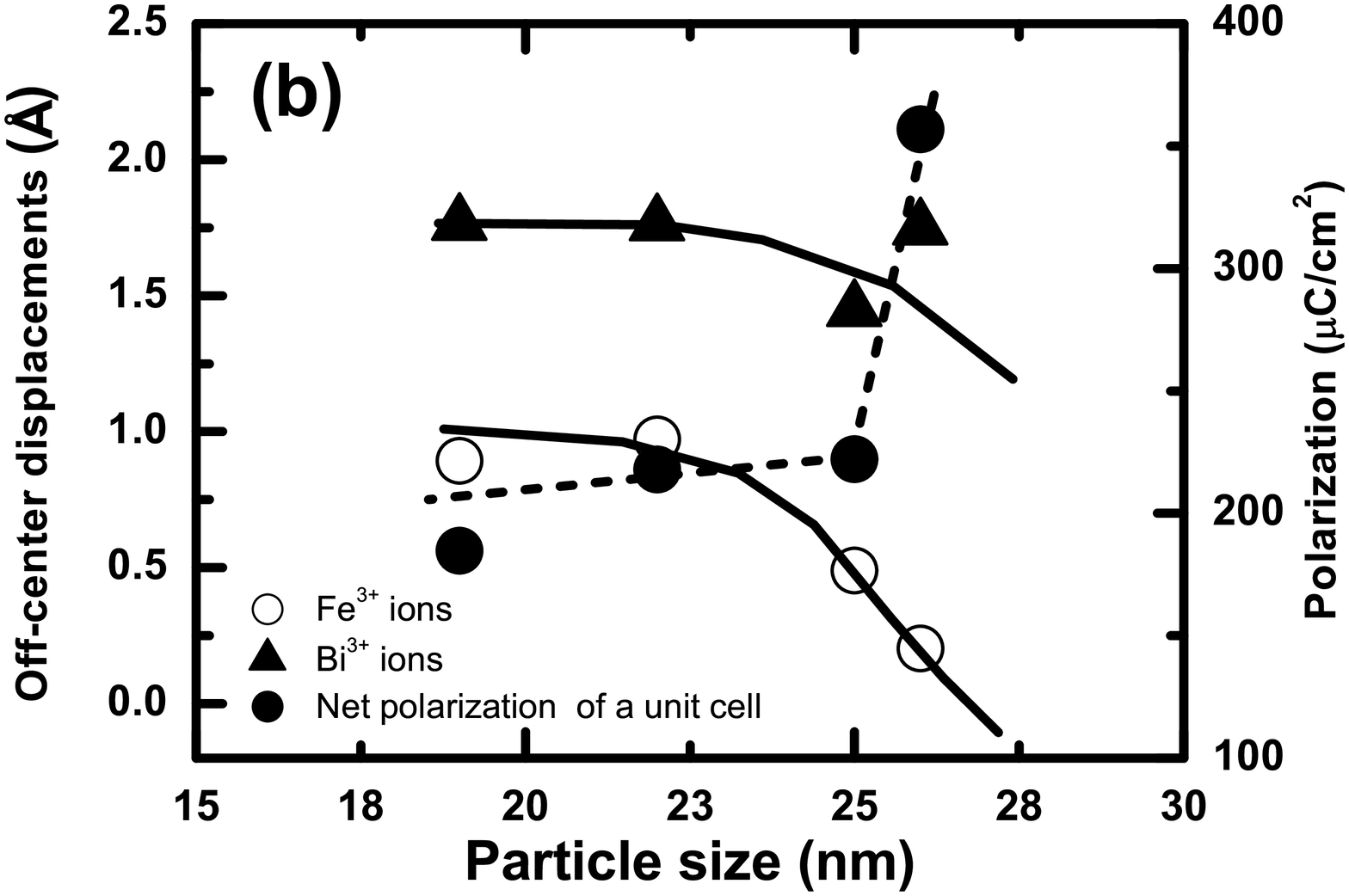}}
\end{center}
\caption{(color online) (a) The experimental and Rietveld refined x-ray diffraction patterns for a representative case ($\sim$22 nm particle); inset shows the corresponding crystallographic structure (rhombohedral with R3c space group); big, medium, and small spheres represent Bi, Fe, and O ions, respectively; (b) variation of off-center displacements of Bi and Fe ions as a function of particle size in the nanoscale regime; variation of net unit cell polarization P$_S$ along the [111]$_{rh}$ $\parallel$ [001]$_{hex}$ with particle size is also shown. }
\end{figure}

The XRD patterns have been refined by Fullprof (ver 2.3) to find out the dependence of non-centrosymmetry of the crystallographic structure on particle size. In BiFeO$_3$, the ferroelectric polarization develops primarily because of covalency in Bi-O bonds resulting from Bi$^{3+}$ 6s$^2$ lone pair and consequent off-center displacement of Bi$^{3+}$ ions with respect to the near-neighbor oxygen network. Interestingly, the Fe$^{3+}$ ions too occupy non-centrosymmetric positions in spite of Jahn-Teller distortion within Fe$^{3+}$O$_6$ octahedra. In Fig. 3a, we show the experimental and refined x-ray diffraction patterns for a representative case ($\sim$22 nm particle). Following points could be noted in the evolution of the patterns: (i) space group changes from R$\bar{3}$c in bulk to lower symmetric R3c in the nanoscale samples; (ii) the strain increases with the decrease in particle size; (iii) the lattice parameters and volume change systematically with particle size. In Table-I, we provide the list of structural parameters as a function of particle size: (i) lattice parameters, (ii) atom positions, (iii) bond lengths and angles within a unit cell, and (iv) lattice strain. The weighted reliability factor R$_{wp}$ which defines the goodness of fit between experimental and refined XRD patterns is also shown.

We estimated the off-center displacements of Fe$^{3+}$ and Bi$^{3+}$ ions within a unit cell of BiFeO$_3$ from the atom position data. In bulk sintered pellet, the $\delta_{Bi}$ and $\delta_{Fe}$ are $\sim$1.235 $\AA$ and $\sim$0.229 $\AA$, respectively. Remarkably, both  $\delta_{Bi}$ and $\delta_{Fe}$ exhibit monotonic rise as the particle size is reduced (Fig. 3b).\cite{Selbach} Interestingly, in contrast to the results in Ref. 8, we found that while  $\delta_{Fe}$ results in polarization along [111]$_{rh}$ $\parallel$ [001]$_{hex}$ axis, $\delta_{Bi}$ is oriented in a different direction which gives polarization along all three directions. Using the  $\delta_{Fe}$ and the component of $\delta_{Bi}$ along [111]$_{rh}$ $\parallel$ [001]$_{hex}$, we estimated the polarization P$_S$ in a unit cell following the procedure described in Ref. 9. P$_S$ is plotted in Fig. 3b as a function particle size. It shows that the resultant P$_S$ of a unit cell increases substantially in nanoscale regime around $\sim$28 nm. It decreases with further decrease in particle size and eventually stabilizes at a value which is much higher than what has been observed in bulk sample \cite{Teague} or even in thin films. \cite{Wang} $\textit{This is the central result of this paper}$. Of course, the large unit cell polarization may not yield large ferroelectric polarization for the entire nanoparticle. This is because of the depolarization originated from surface effects (dead layer) of a nanoparticle\cite{Petkov}. It is necessary to design an appropriate electrode-nanoparticle architecture to cancel out the depolarizing field and thus observe large polarization even across an entire nano-sized particle. 

\section{Conclusion}
We observe significant improvement in unit cell polarization in nanoscale BiFeO$_3$. Both the off-center displacements of Bi$^{3+}$ and Fe$^{3+}$ ions increase monotonically with the decrease in particle size. The nanosized particles also exhibit intrinsic ferromagnetic order. Large magnetostriction resulting from ferromagnetism, possibly, gives rise to striction mediated enhanced ferroelectric polarization in a unit cell. Large ferroelectric polarization and ferromagnetism in nanosized particles will render nanoscale BiFeO$_3$ even more useful than the bulk system for magnetoelectric device applications. 

\textbf{Acknowledgments.}
This work is supported by a Networked research program of CSIR "Nanostructured Advanced Materials" (NWP051). The authors thank J. Ghosh for x-ray diffraction measurements and P.A. Joy for magnetic measurements.

\end{document}